\newcommand{\HII}{\mbox{H\hspace{0.2em}{\scriptsize II}}}

\newcommand{\CII}{\mbox{[C\hspace{0.2em}{\scriptsize II}]}}
\newcommand{\al}{\mbox{$^{26}$\hspace{-0.2em}Al}}
\newcommand{\Msol}{M_{\sun}}

\newcommand{\gray}{\mbox{$\gamma$-ray}}

\newcommand{\um}{\mbox{$\mu$m}}

\def\MeV{\mbox{Me\hspace{-0.1em}V}}

\def\deg{\hbox{$^\circ$}}
\def\sun{\hbox{$\odot$}}

\def\apj{ApJ}

\def   \ni {\noindent}

\def   \ssk {\vskip  5truept}

\def   \bsk {\vskip 15truept}

\def   \newpage {\vfill\eject}
\def   \newline {\hfil\break}

\documentstyle[epsfig]{article}
\begin{document}
\hsize 5truein
\vsize 8truein
\font\abstract=cmr8
\font\keywords=cmr8
\font\caption=cmr8
\font\references=cmr8
\font\text=cmr10
\font\affiliation=cmssi10
\font\author=cmss10
\font\mc=cmss8
\font\title=cmssbx10 scaled\magstep2
\font\alcit=cmti7 scaled\magstephalf
\font\alcin=cmr6
\font\ita=cmti8
\font\mma=cmr8
\def\ref{\par\noindent\hangindent 15pt}
\null

\title{\ni On the Origin of Galactic \raisebox{1ex}{\small 26}Al}

\bsk \bsk
\author{\ni J.~Kn\"odlseder$^1$}
\bsk
\affiliation{$^1$Centre d'Etude Spatiale des Rayonnements, CNRS/UPS, B.P.~4346,
	         31028 Toulouse Cedex 4, France}
\bsk
\baselineskip = 12pt

\abstract{ABSTRACT \ni 
The comparison of COMPTEL 1.8 \MeV\ \gray\ line data to an 
extended data base of tracer maps revealed an extremely close 
correlation between 1.8 \MeV\ gamma-ray line emission on the one side 
and 53 GHz microwave free-free emission and 158 $\mu$m far-infrared 
line emission on the other side (Kn\"odlseder et al., these proceedings).
We demonstrate that the two tracer maps are correlated to the extreme 
Population I, tracing the distribution of very massive stars in the Galaxy.
This strongly suggests that these stars are also the dominant source 
of galactic \al, making Wolf-Rayet stars and/or core collapse supernovae 
the primary candidates.
}
\bsk
\baselineskip = 12pt

\bsk
\baselineskip = 12pt

\text{
\ni 1. INTRODUCTION
\ssk
\ni
\label{sec:intro}

Convincing proof of ongoing nucleosynthesis in the Galaxy comes from 
the observation of the 1.809 \MeV\ gamma-ray line, emitted during the 
decay of radioactive \al.
Theoretical nucleosynthesis calculations suggest that \al\ can be 
produced during nova and core collapse supernova explosions, and 
during hydrostatic hydrogen burning in Asymptotic Giant Branch (AGB) 
stars and Wolf-Rayet (WR) stars.
Considerable uncertainties in nucleosynthesis modelling, however, 
allows for no definite conclusion on the dominant source based on 
theoretical models (Prantzos \& Diehl 1996).
Observations of the 1.809 \MeV\ \gray\ line have the potential to 
answer this question by determination of the spatial distribution
of this \gray\ line.
In particular, they can set constraining limits on stellar yields 
and therefore can help to improve theoretical nucleosynthesis
calculations.

The imaging telescope COMPTEL aboard the {\em Compton Gamma-Ray 
Observatory} ({\em CGRO}) provided the first map of the sky in the 
light of this \gray\ line and enables now a detailed study of the 
sources of galactic radioactivity.
A recent multi-wavelength correlation study of COMPTEL 1.8 \MeV\ data
demonstrated that tracers of the young stellar population provide
reasonable descriptions of galactic 1.809 \MeV\ emission.
The best correlation was found for maps of 53 GHz microwave free-free 
emission and 158 $\mu$m far-infrared line emission
(Kn\"odlseder et al., these proceedings).
Indeed, within the statistics of the present data, the two maps 
provide an entirely satisfactory fit.
We will discuss in this paper how the close correlation between these 
two maps and the COMPTEL 1.8 \MeV\ data can be understood, and 
interpret the results in terms of the galactic \al\ origin.

\newpage
\bsk
\ni 2. FREE-FREE EMISSION
\ssk
\ni
\label{sec:freefree}

Galactic free-free radiation arises from the acceleration of free 
electrons in the electrostatic field of ionized atoms, hence it 
traces the distribution of ionized gas in the interstellar medium 
(ISM).
Most of the free-free emission originates in the extended low-density 
warm ionized medium (ELDWIM) which is spatially correlated to 
galactic \HII\ regions (Heiles 1994).
This is illustrated in Fig.~1 where the longitude profile of 53 GHz 
free-free radiation is compared to the number distribution of \HII\ 
regions as observed by IRAS (Hughes \& MacLeod 1989).
The free-free profile was obtained from the all-sky map of Bennett et 
al.~(1992), who extracted galactic free-free emission from {\em 
COBE} DMR data.
The two longitude profiles correlate very closely, showing the same 
emission peaks -- the fact that structures in the free-free emission 
profile appear broader than in the \HII\ distribution reflects the
poor angular resolution of $7\deg$ (FWHM) of the DMR instrument.

\begin{figure}
\centerline{\psfig{file=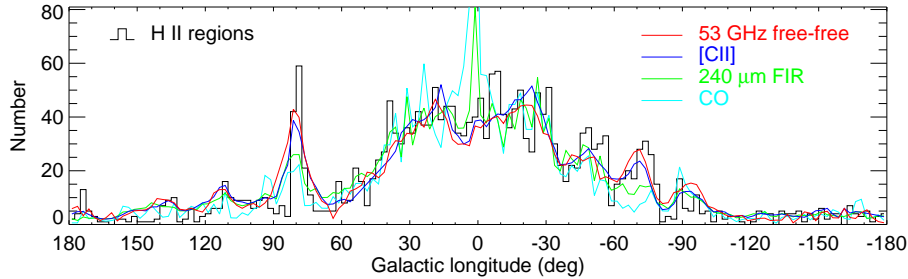, width=12.7cm}}
  \caption{FIGURE 1.
  Longitude profile of IRAS \HII\ regions, 
  53 GHz free-free radiation, 
  158 $\mu$m C$^+$ line emission, 
  240 $\mu$m continuum FIR emission,
  and CO line emission.
  } 
\end{figure}

\HII\ regions are powered by the Lyman continuum (Lyc) photons of 
very massive ($M_{\mathrm{i}}>20\Msol$) stars (e.g.~Abbott 1982).
The ionization source of the ELDWIM component is less clear 
(Heiles et al.~1996) although, due to the power requirement,
very massive stars seem again the most plausible sources.
The close correlation between free-free emission and \HII\ regions 
strongly support this hypothesis (cf.~Fig.~1).
Particular signatures of this extreme Population I are the pronounced 
peaks in Cygnus ($l\approx80\deg$) and Carina ($l\approx-70\deg$) and 
a small dip towards the galactic center region (which can also be 
seen as two broad bumps at $l\sim\pm20\deg$ due to the ``molecular 
ring'' structure at a galactocentric distance of 3-4 kpc).
Other Population I tracers, like far-infrared (FIR) continuum or CO 
line emission, which do not particularly trace the very massive 
component, do not show these signatures.
In these maps the Cygnus feature is much less pronounced, the Carina 
peak is absent, and a pronounced emission peak is seen towards the 
galactic center instead of a dip.
Additionally, these maps show about $10-20\%$ more emission at positive 
longitudes, while the free-free map provides roughly equal 
intensities in both hemispheres.
The COMPTEL 1.8 \MeV\ data show the same characteristics as the 
extreme Population I tracers, while the FIR or CO maps leave 
significant residuals along the galactic plane (Kn\"odlseder et al., 
these proceedings).
We therefore interpret the close correlation between 1.809 \MeV\ line 
and 53 GHz free-free emission as a clear sign that the galactic \al\ 
distribution is correlated to that of very massive stars.

Indeed, Kn\"odlseder (1998) has demonstrated that the free-free 
intensity is in good approximation proportional to the column density 
of ionizing stars.
Since, the 1.809 \MeV\ intensity is proportional to the \al\ column 
density, the observed correlation implies a direct proportionality 
between the \al\ and ionizing star column densities.
Using the fitted proportionality factors derived by Kn\"odlseder et al.~and 
Bloemen et al.~(these proceedings), Kn\"odlseder (1998) obtained 
$Y_{26}^{\mathrm{O7V}} = (1.0 \pm 0.3) \times 10^{-4} \Msol$ 
as typical \al\ yield for an O7 V star (assuming a Lyc luminosity of 
$\log Q_{0}^{\mathrm{O7V}} = 49.05$ ph s$^{-1}$).
Using an estimate for the galactic Lyc luminosity of 
$Q=3.5\times10^{53}$ ph s$^{-1}$ (Bennett et al.~1994), this converts 
into a galactic \al\ mass of $3.1\pm0.9\Msol$.

\bsk
\ni 3. 158 \um\ \CII\ LINE-EMISSION
\ssk
\ni
\label{sec:cii}

\begin{figure}
\centerline{\psfig{file=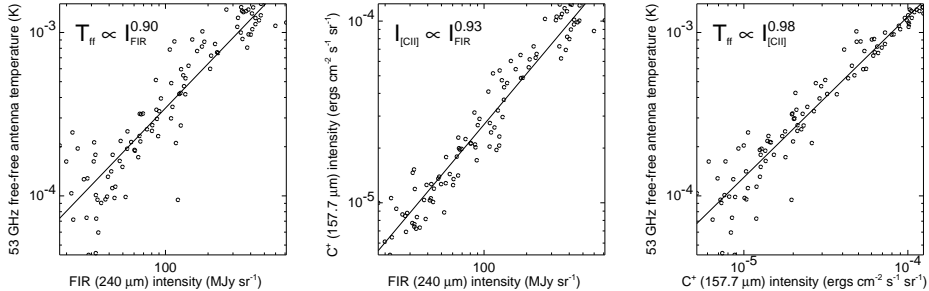, width=12.7cm}}
  \caption{FIGURE 2.
  Correlations between the intensity of free-free radiation, \CII\, 
  and 240 $\mu$m FIR emission in the galactic plane.
  } 
\end{figure}

The 158 \um\ line arises from the fine-structure ground state 
transition of C$^+$, and is considered as the dominant coolant of 
interstellar gas (Tielens \& Hollenbach 1985).
The dominant source of the diffuse galactic 158 \um\ line emission, 
which has recently been mapped by the FIRAS instrument aboard {\em COBE}
(Wright et al.~1991), is still under debate, but there is evidence that 
it could primarily arise from the ELDWIM (Heiles 1994).
This is supported by the longitude profile in Fig.~1 which shows a 
close correlation between free-free and \CII\ emission, and the 
distribution of \HII\ regions.
It can also be seen from the intensity correlations shown in Fig.~2 
which suggest a linear proportionality between free-free and \CII\ 
emission, but smaller slopes for the correlation with FIR continuum 
emission (even flatter slopes between $0.70-0.75$ are obtained for the
correlation with CO emission).
Indeed, as for free-free emission, the \CII\ intensity from the 
ELDWIM component is proportional to the column density of ionizing 
radiation, hence it is also an excellent tracer of the very massive 
star population.
Therefore, the close correlation found between the 1.809 \MeV\ and \CII\ 
line emissions support our hypothesis that the galactic \al\ 
distribution is correlated to very massive stars.

\bsk
\ni 4. DISCUSSION
\ssk
\ni

The close correlations of 53 GHz free-free emission and 158 
\um\ \CII\ radiation to COMPTEL 1.8 \MeV\ data suggest that the 
galactic \al\ distribution is associated to the ELDWIM component.
To maintain its ionization, the ELDWIM must be irradiated by Lyc 
photons which can only be provided in sufficient amounts by very 
massive stars.
This strongly suggests that \al\ is itself correlated to the very 
massive star population, which is therefore the most plausible origin 
of this radioactive isotope.
Indeed, the short lifetime of $\tau_{26}\sim10^6$ years guarantees 
that \al\ is never found too far away from its origin, allowing for 
the link between 1.809 \MeV\ emission and the sources of \al.
Thus, core-collapse supernovae ($M_{\mathrm{i}}>8\Msol$) 
and/or Wolf-Rayet stars ($M_{\mathrm{i}}>25-35\Msol$) should be 
considered as primary candidate sources of galactic \al.
AGB stars, even massive ones, and novae, which are not correlated to 
the ionizing stellar population, are probably not prolific sources 
of \al.
This is supported by the fact that FIR continuum or CO line emission, 
which also trace the formation of non-ionizing stars, do not provide 
a globally satisfactory correlation to 1.8 \MeV\ data.
In particular, tracers of the old stellar population show no 
correlation to 1.809 \MeV\ emission (Kn\"odlseder et al., these 
proceedings).

\bsk
\baselineskip = 12pt
{\abstract \ni ACKNOWLEDGMENTS
The COMPTEL project is supported by the German government through
DARA grant 50 QV 90968, by NASA under contract NAS5-26645, and by
the Netherlands Organisation for Scientific Research NWO.
The {\em COBE} datasets were developed by the NASA Goddard Space Flight 
Center under the guidance of the {\em COBE}  Science Working Group and were 
provided by the NSSDC.
The research project has also benefit from financial support 
provided by the Cit\'e de l'Espace in Toulouse. 
}
\bsk
\baselineskip = 12pt

{\references \ni REFERENCES
\ssk

\ref Abbott, D.~C.~1982, ApJ, 263, 723
\ref Bennett, C.~L. et al.~1994, ApJ, 434, 587
\ref Bennett, C.~L., Smoot, G.~F., Hinshaw, G., et al.~1992, 
     \apj, 396, L7
\ref Heiles, C.~1994, ApJ, 436, 720
\ref Heiles, C., et al.~1996, ApJ, 462, 326
\ref Hughes, V.~A.~\& MacLeod, G.~C.~1989, AJ, 97, 786
\ref Kn\"odlseder, J.~1998, ApJ, in press
\ref Prantzos, N.~\& Diehl, R.~1996, Phys.~Rep., 267, 1
\ref Tielens, A.~G.~G.~M.~\& Hollenbach, D.~1985, ApJ, 291, 722
\ref Wright, E.~L., et al.~1991, ApJ, 381, 200

}                      

\end{document}